\documentclass[a4paper,12pt]{article}
\usepackage{amsmath,amssymb}
\usepackage[nosort]{cite}

\begin{document}
\begin{titlepage}
\pagestyle{empty}

\begin{center}

\begin{flushright}
{\sc\footnotesize hep-th/0604069}\\
{\sc\footnotesize NORDITA-2006-11}\\
\end{flushright}
\vspace*{100pt}

\begin{center}
{\Large \bf {A Universality Test of the Quantum String Bethe Ansatz}}
\end{center}

\vspace{26pt}

{L. Freyhult\footnote{freyhult@nordita.dk} and C. Kristjansen\footnote{kristjan@nbi.dk}}
\\
\vspace{6pt}
{ NORDITA,} \\
{ Blegdamsvej 17, DK-2100 Copenhagen \O}\\

\vspace{20pt}
\end{center}

\vspace{5mm}
\begin{center}
{\large \bf Abstract}
\end{center}
\noindent
We show that the quantum corrected string Bethe ansatz passes an important
universality test by demonstrating that it correctly
incorporates the non-analytical terms in the
string sigma model one-loop correction for rational three-spin strings with
two out of the three spins identical. Subsequently, we use the quantum
corrected string Bethe ansatz to predict the exact form of the non-analytic
terms for the generic rational three-spin string.

\end{titlepage}
\newpage

\pagestyle{plain}

\setcounter{page}{1}

\section{Introduction \label{introduction}}

The discovery of integrable structures governing the spectrum of ${\cal N}=4$ 
SYM~\cite{Minahan:2002ve,Beisert:2003tq,Beisert:2003yb} as well as that of type
IIB superstrings on $AdS_5\times S^5$~\cite{Mandal:2002fs,Bena:2003wd}
has shifted the focus away from the dilatation operator of the gauge theory respectively the string
Hamiltonian towards that of the S-matrices
of the underlying integrable models~\cite{Staudacher:2004tk}.

On the ${\cal N}=4$ SYM side the relevant $S$-matrix has been determined 
in perturbation theory to three-loop order in certain 
sectors~\cite{Staudacher:2004tk}
and a conjecture for an all sector, all loop $S$-matrix 
exists~\cite{Beisert:2005fw}. In addition, S-matrix techniques have
proved useful for the study of various generalizations of the
gauge theory~\cite{Beisert:2005wv,Freyhult:2005ws}.

Due to our current inability to quantize the string theory on $AdS_5\times S^5$
the S-matrix underlying the integrable structures on the string theory side
is less well understood. Based on input from pp-wave physics as well as various
semi-classical analyses it has been
 proposed that the string theory $S$-matrix 
differs
from the gauge theory one only by a phase factor, expressible in terms of certain 
conserved charges~\cite{Arutyunov:2004vx,Staudacher:2004tk,Beisert:2005fw}, 
see also~\cite{Beisert:2005tm,Janik:2006dc}. 
\begin{equation}
S^{string}(p_k,p_j)=e^{2 i\theta(p_k,p_j)}S^{gauge}(p_k,p_j).
\end{equation}
The most general possible form of the phase consistent with integrability
and respecting the symmetries of
${\cal N}=4$ SYM 
reads~\cite{Beisert:2005wv} 
\begin{equation}
\theta(p_k,p_j)=\sum_{r=2}^{\infty}\sum_{s=r+1}^{\infty}
c_{r,s}(\lambda)\left(\frac{\lambda}{16 \pi^2}\right)^{\frac{r+s-1}{2}}
\left(q_r(p_k)q_s(p_j)-q_s(p_j)q_r(p_k)\right),
\label{correction}
\end{equation}
where $c_{r,s}=0$ unless $r+s$ is odd. Here,
$q_r(p_k)$ is the contribution to the $r$'th conserved charge
from the $k$'th excitation and $\lambda$ is the 't Hooft coupling constant.

With $c_{r,s}=\delta_{s,r+1}$ the above string $S$-matrix gives rise to
the correct continuum Bethe equations describing the motion of classical
strings on $AdS_5\times S^5$~\cite{Kazakov:2004qf,
Kazakov:2004nh,Beisert:2004ag,Schafer-Nameki:2004ik,
Beisert:2005bm,Beisert:2005di} and describes correctly
strings in the BMN-limit~\cite{Berenstein:2002jq} and the near 
BMN-limit~\cite{Callan:2003xr}.
In particular, this string $S$-matrix accounts for the observed three-loop
discrepancy between gauge theory and string 
theory~\cite{Callan:2004ev,Callan:2004uv,Serban:2004jf}.
In all of these instances the string energies are 
analytic in $\lambda$. String sigma model one-loop corrections, however,
lead to half-integer powers of $\lambda$ appearing in the expressions for
energies of spinning 
strings~\cite{Schafer-nameki:2005tn,Schafer-nameki:2005is,Beisert:2005cw}. 
It was suggested
that these could be accommodated in the string Bethe ansatz by assuming
$c_{r,s}$  to have an expansion in half-integer 
inverse powers of $\lambda$, i.e.~\cite{Beisert:2005cw}
\begin{equation}
\label{largelambda}
c_{r,s}(\lambda)=\delta_{s,r+1}+\delta c_{r,s}\,\frac{1}{\sqrt{\lambda}}+
{\cal O}\left(\frac{1}{\lambda}\right).
\end{equation}
Here, the first $1/\sqrt{\lambda}$ correction term should be enough to 
account for the string sigma model one-loop correction.
This idea was tested for 
strings dual to gauge theory operators in respectively
the ${\mathfrak{su}}(2)$ sector
at half-filling and the ${\mathfrak{sl}}(2)$ 
sector~\cite{Beisert:2005cw,Hernandez:2006tk}. Whereas the 
$\mathfrak{su}(2)$ case did
not fix the corrections to the coefficients $c_{r,s}$ uniquely, the
study of the $\mathfrak{sl}(2)$ sector did and lead to a conjecture for 
$\delta c_{r,s}$~\cite{Hernandez:2006tk}. As the quantum string Bethe ansatz 
conjecture implies
that the dressing factor should be universal~\cite{Beisert:2005fw}, 
it is important that the 
coefficients $\delta c_{r,s}$ be the same for all sectors. In the present
paper we show that the quantum string Bethe ansatz passes an important
universality test by demonstrating that string sigma model one loop
corrections in the ${\mathfrak{su}}(3)$ sector, i.e.\ for strings with
three large angular momenta on $S^5\subset AdS_5\times S^5$, also uniquely fix the
$\delta c_{r,s}$ and that to values coinciding with
the conjectured ones.  

The paper is organised as follows. In section~\ref{SBA} we formulate the
quantum corrected string Bethe ansatz for strings with three large angular
momenta on $S^5$
and
use it to derive the first non-analytical contributions to 
energies of rational three-spin strings.
Subsequently, in section~\ref{stringtheory} we determine the non-analytical
corrections via a one-loop string sigma model calculation for the case of
a three-spin string with two out of the three spins identical. Matching the
two results we in section~\ref{coeff} uniquely determine the 
coefficients $\delta c_{r,s}$ in the quantum corrected string Bethe ansatz. We
furthermore use these results to predict
the string sigma model one-loop corrections
for rational three spin strings with arbitrary values of the three angular 
momenta and the
two possible winding numbers. Finally,
section~\ref{conclusion} contains our conclusions.

\section{Non-analytic corrections from the string Bethe ansatz
\label{SBA}}

The conjectured string Bethe ansatz~\cite{Beisert:2005fw} 
does not in general allow one to treat the $\mathfrak{su}(3)$ sector as 
closed but rather demands that one considers the
larger $\mathfrak{su}(2|3)$ sub-sector. However, as we are interested in
studying spinning strings we will always be working in
the thermodynamical limit 
where the $\mathfrak{su}(3)$ sector is effectively 
closed~\cite{Minahan:2004ds}.

In the thus effectively closed $\mathfrak{su}(3)$ sector one has two types
of Bethe roots with filling fractions $\alpha$ and $\beta$ related to
the angular momentum labels of the string by
\begin{equation}
(J_1,J_2,J_3)=J(1-\alpha,\alpha-\beta,\beta),
\end{equation}
where $J$ is the length of the spin chain. The thermodynamical limit
of the proposed string Bethe equations read
\begin{eqnarray}\label{eq:BE}
V_1(\varphi)
&=&2-\hspace{-0.47cm}\int\frac{\rho_1(\varphi')}{\varphi-\varphi'}-\int\frac{\rho_2(\varphi')}{\varphi-\varphi'},\\
V_2(\varphi)&=&2-\hspace{-0.47cm}\int\frac{\rho_2(\varphi')}{\varphi-\varphi'}-\int\frac{\rho_1(\varphi')}{\varphi-\varphi'},
\end{eqnarray}
with $\rho_1(\varphi)$ and $\rho_2(\varphi)$ 
being the densities of Bethe roots 
with filling fractions $\alpha$ and $\beta$ respectively.
In the case of rational three spin strings the potentials are
\begin{eqnarray}
&&V_1(\varphi)=2\pi n+\frac{1}{\sqrt{\varphi^2-4\omega^2}}-\sum_{r,s}\omega^{r+s-1}c_{r,s}(\lambda)\left(q_r(\varphi)Q_s-Q_rq_s(\varphi)\right),\\
&&V_2(\varphi)=2\pi m,
\end{eqnarray}
where $m$ and $n$ are integers and
where the charges are given by
\begin{equation}
q_r(\varphi)=\frac{1}{\sqrt{\varphi^2-4\omega^2}}\frac{1}{\left(\frac{1}{2}\varphi+\frac{1}{2}\sqrt{\varphi^2-4\omega^2}\right)^{r-1}},\quad Q_r=\int d\varphi'\rho_1(\varphi')q_r(\varphi').
\end{equation}
Here we have used the notation
\begin{equation}
\omega=\frac{1}{4\pi\mathcal{J}}=\frac{\sqrt{\lambda}}{4\pi J}.
\end{equation}
The string Bethe equations have to be supplemented by the level
matching condition
\begin{equation}
Q_1 = 2 p \pi,\hspace{0.7cm} p\in \mathbb{Z}. \label{level}
\end{equation}
We are interested in determining the leading 
non-analytical contributions to the string energy, i.e.\
\begin{equation}
\delta {\mathcal E}_{non-analytic}= (Q_2)_{non-analytic}\,. 
\end{equation}
It is easy to see that
with the ansatz~(\ref{largelambda}) for the coefficients $c_{r,s}$ 
the corrections $\delta c_{r,s}$
will
for finite $\omega$ be of 
the same order as the leading finite size effects studied 
in~\cite{Freyhult:2005fn} (and for the $\mathfrak{su}(2)$ and
$\mathfrak{sl}(2)$ sector 
in~\cite{Lubcke:2004dg,Beisert:2005mq,Hernandez:2005nf}). 
As the leading
finite size corrections are {\it analytical} in $\lambda$ we, however, do not
need to worry about mixing of the two effects and can safely ignore the
finite size corrections.

Using a similar trick as in \cite{Freyhult:2005fn} we 
can turn the above relations into a quadratic and a cubic equation. 
Let us introduce the resolvents
\begin{equation}
G_1(\varphi)=\int\frac{\rho_1(\varphi')}{\varphi-\varphi'}, \quad G_2(\varphi)=\int\frac{\rho_2(\varphi')}{\varphi-\varphi'}.
\end{equation}
which have the following behaviour as $\varphi\rightarrow \infty$ 
\begin{equation}
G_1(\varphi)\to\frac{\alpha}{\varphi},
\quad G_2(\varphi)\to\frac{\beta}{\varphi}.
\end{equation}
The quadratic and cubic equations can then be written as
\begin{eqnarray}
&&\hspace{-1cm}\nonumber\label{eq:quadratic}G_1^2(\varphi)+G_2^2(\varphi)-G_1(\varphi)G_2(\varphi)-V_2(\varphi)G_2(\varphi)-V_1(\varphi)G_1(\varphi)\\
&&\hspace{-1cm}+\mbox{Res}_{x=0}\left(\frac{G_1(x)V_1(x)}{\varphi-x}\right)=0,\\
&&\hspace{-1cm}\nonumber \label{eq:cubic}G_2^2(\varphi)G_1(\varphi)-G_1^2(\varphi)G_2(\varphi)-V_2(\varphi)G_2^2(\varphi)+V_2^2(\varphi)G_2(\varphi)+V_1(\varphi)G_1^2(\varphi)-V_1^2(\varphi)G_1(\varphi)\\
&&\hspace{-1cm}-\mbox{Res}_{x=0}\left(\frac{G_1^2(x)V_1(x)}{\varphi-x}\right)+\mbox{Res}_{x=0}\left(\frac{G_1(x)V^2_1(x)}{\varphi-x}\right)=0.
\end{eqnarray}
Let us denote the negative moments of the root distributions as 
${\mathcal Q_i}$ and ${\mathcal P_i}$, i.e.\ 
\begin{equation}
G_1(\varphi)=-\sum_{i=0}^{\infty}{\mathcal Q}_{i+1}\,\varphi^i,
\quad G_2(\varphi)=-\sum_{i=0}^{\infty}{\mathcal P}_{i+1}\,\varphi^i.
\end{equation}
{}From simply integrating over $\varphi$ in (\ref{eq:BE}) we find
\begin{equation}\label{eq:Momcond}
Q_1=2\pi (n\alpha+ m\beta). 
\end{equation}
Expanding the cubic and quadratic equation for large and small values of $\varphi$ allows us to determine the resolvents perturbatively in the coupling, $\omega$. 
The cubic equation at infinity gives $Q_2$ in terms of the moments 
${\mathcal Q}_i$ and ${\mathcal P}_i$. (The quadratic equation at infinity 
is redundant as it is equivalent to the momentum condition.)
The moments are obtained 
by first eliminating $Q_1$ using the momentum
constraint and next using the quadratic and cubic equation 
interchangeably to determine ${\mathcal P_i}$ and 
${\mathcal Q}_i$. The algorithm obviously works for any values of 
$\alpha$, $\beta$, $m$ and $n$, i.e.\ for any rational three-spin 
string, but for simplicity we shall present the result only for
the case where two out of the three angular momenta of the string 
are equal\footnote{For an example of the complexity of the general
result, see section~\ref{coeff}.}. Thus, let us specialize to the 
case\footnote{This choice allows us to get the result for all 
possible three-spin strings with two out of three angular momenta
identical, see detailed discussion in\cite{Freyhult:2005fn}.}
\begin{equation}
\beta=\frac{\alpha}{2},\quad \,\,\, n=-\frac{m}{2}\equiv k,
\end{equation}
where $k$ is now the single remaining winding number of the string.
Using the above discussed
 expansions of (\ref{eq:quadratic}) and (\ref{eq:cubic}) we find the 
following
non-analytic contribution to the energy\footnote{The calculation can be
taken much further but for simplicity we limit ourselves to presenting
only the first
three terms.}
\begin{eqnarray}\label{deltaEstringBA}
&&\hspace{-1cm}\nonumber \delta \mathcal{E}_{non-analytic}=\frac{1}{16\mathcal{J}^5}\delta c_{2,3}k^6\alpha^2(3\alpha-2)\\
&&\hspace{-1cm}\nonumber+\frac{1}{64\mathcal{J}^7}k^8\alpha^2\left((10-23\alpha+8\alpha^2)\delta c_{2,3}+(2-7\alpha+6\alpha^2)\delta c_{3,4}-2(1-5\alpha+5\alpha^2)\delta c_{2,5}\right)\\
&&\hspace{-1cm}\nonumber+\frac{1}{256\mathcal{J}^9}k^{10}\alpha^2\left(\delta c_{2,3}(-42+131\alpha-64\alpha^2-24\alpha^3)+\delta c_{3,4}(-14+65\alpha-88\alpha^2+31\alpha^3)\right.\\
&&\hspace{1.2cm}\nonumber\left.+\delta c_{2,5}(14-94\alpha+154\alpha^2-63\alpha^3)+\delta c_{2,7}(-2+21\alpha-56\alpha^2+42\alpha^3)\right.\\
&&\hspace{1.2cm}\left.+\delta c_{3,6}(2-15\alpha+32\alpha^2-21\alpha^3)+\delta c_{4,5}(-2+14\alpha-30\alpha^2+20\alpha^3)\right)+\dots
\end{eqnarray}

\section{Non-analytic corrections from a string sigma model one loop
calculation
\label{stringtheory}}

The string sigma model one-loop correction has only been explicitly worked
out for two classes of classical string solutions. One class 
is rational two-spin
strings with one large momentum on $S^5$ and one on $AdS_5$,
dual to operators in the ${\mathfrak{sl}}(2)$ sector of 
the gauge theory~\cite{Park:2005ji}. The other class is rational three spin
strings with three large angular momenta on $S^5$, two of them being
identical~\cite{Frolov:2003qc,Frolov:2003tu,Frolov:2004bh}. 
(For partial results on
the generic three-spin case, see~\cite{Arutyunov:2003za,Fuji:2005ry}.)
 In the present paper we shall consider the latter case denoting the
three spins as $(\mathcal{J}_1,\mathcal{J}_2,\mathcal{J}_2)$ and the
sum of these as $\mathcal{J}$.
(When $\mathcal{J}_1$ vanishes this
case reduces to two-spin strings dual to operators in the
$\mathfrak{su}(2)$ sector at half-filling.)
The one-loop energy
receives contributions from fluctuations in the bosonic (B) as well as
the  fermionic (F)
fields both in the $S^5$ and in the $AdS_5$ part of the space and
takes the form~\cite{Frolov:2003tu,Frolov:2004bh}
\begin{equation}\label{deltaE}
\delta E=\frac{1}{2\kappa}\left(\sum_{n\in \mathbb{Z}}\omega_n^B-\sum_{r\in \mathbb{Z}+1/2}\omega_r^F\right),
\end{equation}
where $\kappa$ relates world-sheet and the $AdS_5$ time as $t=\kappa\tau$.
Here $\omega$ is the absolute value of the 
frequencies\footnote{This is not true in the general case, however as
  long as 
half of the frequencies are negative and the other half positive this
holds. 
See \cite{Blau:2003rt,Park:2005ji} for details.}.
The contributions to $\omega^B$ from fluctuations in the $S^5$
directions 
are given by the roots of the polynomial
\begin{eqnarray}
\lefteqn{\nonumber B_8(\Omega)=\Omega^4+\Omega^3 \left(-8k^2-4 n^2+20k^2 q-8 \kappa ^2\right)+\Omega^2(16k^4+8 k^2n^2+6 n^4-80k^4 q}\\
&&\nonumber-36k^2n^2 q+96 k^4q^2+32k^2 \kappa ^2+16 n^2 \kappa ^2-80 k^2q \kappa ^2+16 \kappa ^4)+
\Omega (-32 k^4n^2+8 k^2n^4\\
&&\nonumber-4 n^6+96 k^4n^2 q+12k^2 n^4 q-96 k^4n^2 q^2-32 k^2n^2 \kappa ^2-8 n^4 \kappa ^2+48k^2 n^2 q \kappa ^2)\\
&&+16 k^4n^4-8k^2 n^6+n^8-16 k^4n^4 q+4 k^2n^6 q,
\end{eqnarray}
where $\Omega=\omega^2$ and
\begin{equation}
\kappa=\sqrt{\nu^2+2q},\quad \,\,\,q=\frac{\alpha\mathcal{J}}{\sqrt{\nu^2+1}},
\end{equation}
and $\nu$ is given by
\begin{equation}
\left(\nu-\mathcal{J}(1-\alpha)\right)\sqrt{1+\nu^2}-\alpha\mathcal{J}\nu=0.
\end{equation}
We have introduced $\alpha$ to denote the fraction $2\mathcal{J}_2/\mathcal{J}$.
The strings we consider are localized at the center of $AdS_5$ and 
the bosonic frequencies associated with fluctuations in that part of the space therefore take the simple form
\begin{equation}
\omega_{AdS}=\sqrt{n^2+\kappa^2}.
\end{equation}
The fermionic frequencies mix fluctuations in $S^5$ and $AdS_5$ and are given by the roots of the polynomial
\begin{eqnarray}
\lefteqn{\nonumber F_8(\Omega)=2\Omega^4+\Omega^3(-8k^2-12\kappa^2
-8r^2+20k^2q)}\\
&&\nonumber+\Omega^2(12k^4+28k^2\kappa^2+18\kappa^4+8k^2r^2+28\kappa^2r^2+12r^4 \\
&&\nonumber
-52k^4q-64k^2\kappa^2q-36k^2r^2q+59k^4q^2)\\
&&\nonumber+\Omega(-8k^6-20k^4\kappa^2-20k^2\kappa^4-8\kappa^6+8k^4r^2+8k^2r^2\kappa^2-20\kappa^4r^2+8k^2r^4-20\kappa^2r^4\\
&&\nonumber-8r^6+44k^6q+80k^4\kappa^2q+44k^2\kappa^4q-24k^4r^2q+32k^2\kappa^2r^2q+12k^2r^4q-78k^6q^2\\
&&\nonumber-79k^4\kappa^2q^2+2k^4r^2q^2+45k^6q^3)
\\&&\nonumber 
+2k^8+4k^6\kappa^2+2k^4\kappa^4-8k^6r^2-4k^4\kappa^2r^2-4k^2\kappa^4r^2
 +12k^4r^4-4k^2\kappa^2r^4+2\kappa^4r^4\\
&&\nonumber
-8k^2r^6+4\kappa^2r^6+2r^8-12k^8q-16k^6\kappa^2q-4k^4\kappa^4q+28k^6qr^2
+16k^4\kappa^2r^2q\\
&&\nonumber
+4k^2\kappa^4r^2q-20k^4r^4q+4k^2r^6q+27k^8q^2+21k^6\kappa^2q^2+2k^4\kappa^4q^2
-30k^6r^2q^2\\
&&-11k^4\kappa^2r^2q^2+11k^4r^4q^2-27k^8q^3-9k^6\kappa^2q^3+9k^6r^2q^3+81/8k^8q^4.
\end{eqnarray}
 The expression for the one-loop energy shift (\ref{deltaE}) is convergent, the constituent bosonic and fermionic sums are, however, divergent. 
Rearranging the sums taking into account that they individually diverge (\ref{deltaE}) can be rewritten as in \cite{Frolov:2004bh}
\begin{equation}
\delta E=\frac{1}{2\kappa}\left[2+\sum_{n\in\mathbb{Z}}\left(\omega^B_n-\frac{1}{2}\left(\omega_{n+1/2}^F+\omega_{n-1/2}^F\right)\right)\right].
\end{equation}
The sum is again convergent. However, the individual terms in the $1/\mathcal{J}$-expansion are divergent. A way to get around an artificial regularisation of the sum is to split the one-loop correction as 
follows~\cite{Beisert:2005cw,Schafer-nameki:2006gk}
\begin{equation}
\delta E=\sum_{n=-\infty}^\infty e^{sum}_{reg}(n)+
{\mathcal J}\int_{-\infty}^\infty dx \,e^{int}_{reg}(x),
\end{equation}
where the last part is equal to the divergent part of the expanded sum
and 
encodes the non-analytic contribution. 
It can be computed, following the procedure in \cite{Beisert:2005cw} ,by first making the rescaling $n=\mathcal{J}x$, then expanding in $1/\mathcal{J}$ and finally integrating over the regular part of the obtained expression. 
The computations can be simplified setting the winding number, $k$, to one. It is always possible to, in the end, restore it by rescaling the variables as 
\begin{equation}
n\to \frac{n}{k},\quad \nu\to\frac{\nu}{k}\quad \mbox{and}\quad \omega\to\frac{\omega}{k}. 
\end{equation}
Setting $k=1$, rescaling $n$ and expanding we find
\begin{eqnarray}
\lefteqn{
\nonumber e^{int}(x)=-\frac{1}{\mathcal{J}^2}\frac{1}{2 \left(1+x^2\right)^{3/2}}}\\
&&\nonumber-\frac{1}{\mathcal{J}^4}\frac{16 \alpha ^2-x^2 \left(13+12 \alpha -80 \alpha ^2\right)+x^4 \left(52-48 \alpha +64 \alpha ^2\right)+24 x^6 \alpha}{32
x^2 \left(1+x^2\right)^{7/2}}\\
&&\nonumber-\frac{1}{\mathcal{J}^6}\frac{1}{256 x^4 \left(1+x^2\right)^{11/2}}\left(256 \alpha ^2+64 x^2 \alpha ^2(21-\alpha )\right.\\
&&\nonumber\left.+x^4 (151-106 \alpha +2504 \alpha ^2-64 \alpha ^3)-4 x^6 (453-668 \alpha -652 \alpha ^2+312 \alpha ^3)\right.\\
&&\nonumber\left.+4 x^8 (302-687
\alpha+ 940 \alpha ^2-360 \alpha ^3)+16 x^{10} \alpha
 (4+3 \alpha -12 \alpha ^2)-80 x^{12} \alpha\right)
\\
&&\nonumber-\frac{1}{\mathcal{J}^8}\frac{1}{8192 x^6 \left(1+x^2\right)^{15/2}}
(4096 \alpha ^2 (8-6 \alpha +3 \alpha ^2)+512 x^2 \alpha ^2 (463-347 \alpha +166 \alpha ^2)\\
&&\nonumber+256 x^4 \alpha ^2 (2871-2153 \alpha +978 \alpha ^2)\\
&&\nonumber-x^6 (7565-17832 \alpha -1267952 \alpha ^2+993152 \alpha ^3-436224 \alpha ^4)\\
&&\nonumber+8 x^8 (22695-55296 \alpha +198892 \alpha ^2-123632 \alpha ^3+46720 \alpha ^4)\\
&&\nonumber+16 x^{10} (-22695+54845 \alpha +4998 \alpha ^2-25536 \alpha ^3+14928 \alpha ^4)\\
&&\nonumber+32 x^{12} (3026-9319 \alpha +19181 \alpha ^2-13356 \alpha ^3+5040 \alpha ^4)\\
&&\nonumber+96 x^{14} \alpha  (249-190 \alpha -192 \alpha ^2+456 \alpha ^3)\\
&&+128 x^{16} \alpha  (31-126 \alpha +106 \alpha ^2+48 \alpha ^3)+1792 x^{18} \alpha  (1-2 \alpha +2 \alpha ^2))+\dots \label{eint}
\end{eqnarray}
We are able to go a few orders higher in the expansion 
(cf.\ eqn.~(\ref{deltaEstring}))
but we omit
those terms above as the expressions become very involved.
The singular part of the above expression is
\begin{eqnarray}
e^{int}_{sing}(x)&=&-\frac{1}{\mathcal{J}^6}\left(\frac{1}{x^4}\alpha ^2+\frac{1}{4x^2}\alpha(1+\alpha ^2)\right)\\
&&+\frac{1}{\mathcal{J}^8}\left(\frac{1}{2x^6}\alpha^2(8-6\alpha+3\alpha^2)-\frac{1}{16x^4}\alpha^2(17-13\alpha+14\alpha^2)\right. \nonumber \\
&&\left.\hspace{1.0cm}
+\frac{1}{16x^2}\alpha^2(3-\alpha+9\alpha^2)\right)+\dots \nonumber
\end{eqnarray}
Integrating the regular part and restoring the $k$-dependence we find
\begin{eqnarray}\label{deltaEstring}
\lefteqn{\nonumber \mathcal{J}\int_{-\infty}^{\infty}dxe_{int}^{reg}(x)=\frac{1}{\mathcal{J}^5}\frac{1}{3}  k^6\alpha ^2(2-3 \alpha )-\frac{1}{\mathcal{J}^7}
\frac{1}{30}k^8 \alpha ^2 \left(32-79 \alpha +37 \alpha ^2\right)}\\
&&\nonumber+\frac{1}{1680\mathcal{J}^9}k^{10}\alpha^2(1152 - 4989\alpha + 6956\alpha^2 - 3462\alpha^3) \\
&&-\frac{1}{5040 \mathcal{J}^{11}}k^{12} \alpha ^2 (8192-55451 \alpha +134670 \alpha ^2-139535 \alpha ^3+50108 \alpha ^4)+\dots
\end{eqnarray}
We notice that the integral of the two first terms in the 
expansion~(\ref{eint})
vanish 
upon integration. A similar situation is encountered in the case of
strings dual to operators in the $\mathfrak{sl}(2)$ 
sector~\cite{Beisert:2005cw}.

\section{Determination of the coefficients \label{coeff} }
Comparing the two obtained expressions for the non-analytic contribution 
the string energy, (\ref{deltaEstring}) and (\ref{deltaEstringBA}) one
can uniquely determine the coefficients $\delta c_{r,s}$.
This leads to  the following results 
\begin{eqnarray}
&&\nonumber\delta c_{2,3}=-16/3,\\
&&\nonumber\delta c_{3,4}=-48/3,\quad\delta c_{2,5}=-32/15,\\
&&\nonumber\delta c_{4,5}=-96/7,\quad\delta c_{3,6}=-80/21,\quad\delta c_{2,7}=-48/35,\\
&&\delta c_{5,6}=-160/9,\quad\delta c
_{4,7}=-16/3,\quad\delta c_{3,8}=-112/45,\quad\delta c_{2,9}=-64/63,
\end{eqnarray}
which do indeed confirm the conjectured formula~\cite{Hernandez:2006tk}
\begin{equation}\label{coefficients}
\delta c_{r,s}=
\left\{\begin{array}{c}-8\frac{(r-1)(s-1)}{(r+s-2)(s-r)}\quad\mbox{if $r+s$ odd}\\\hspace{1cm}0 \hspace{1.9cm}\mbox{if $r+s$ even,}\end{array}\right.
\end{equation}
where it is understood that $r\geq 2$ and $s\geq r+1$.
Assuming now the conjecture to be true one can, using the results of
section~\ref{SBA}, predict the 
non-analytical contributions to the energies of rational three-spin 
strings with three different angular momenta 
$(\mathcal{J}_1,\mathcal{J}_2,\mathcal{J}_3)=
\mathcal{J}(1-\alpha,\alpha-\beta,\beta)$ and winding numbers $(m,n)$.
(Note, that according to eqns.~(\ref{level}) and~(\ref{eq:Momcond})
the quantity $n\alpha+m\beta$ has to be an integer.) 
The
result for the leading correction reads\footnote{Here again it is 
possible to go to much higher orders.}
\begin{eqnarray}
\lefteqn{
\delta {\mathcal E}_{non-analytical}=} \\
&&-\frac{1}{{\mathcal J}^5}\left\{
\frac{1}{3}\, n^6 (1-\alpha)^3\alpha^3+
\frac{1}{3}\, m^6 (1-\beta)^3 \beta^3\right.
\nonumber\\
&&\,\,\,\,\,\,\,\,\,\,\,+2\, m n^5 (1-\alpha)^3\alpha^2 \beta+
2\,m^5 n(1-\alpha)(1-\beta)^2 \beta^3 \nonumber\\
&&\,\,\,\,\,\,\,\,\,\,
+\frac{1}{3}\, m^2 n^4 (1-\alpha)\beta (3\alpha^2(1-9\beta)+2\beta
+3\alpha^2(-1+5\beta)+2\alpha(-1+6\beta))\nonumber \nonumber \\
&&\,\,\,\,\,\,\,\,\,\,+\frac{4}{3}\, m^3 n^3(1-\alpha) \beta (\beta(1+2\beta)+\alpha^2 \beta (-3+5\beta)+\alpha(-1+3\beta-7\beta^2))\nonumber \\
&&\,\,\,\,\,\,\,\,\,\,\left.+\frac{1}{3}\, m^4 n^2 (1-\alpha) \beta(2\beta (1+6\beta-6\beta^2)+\alpha(-2+3\beta-18\beta^2+15\beta^3))\right\}. 
\nonumber
\end{eqnarray}

\section{Conclusion \label{conclusion}}

We have performed an important universality test of the conjectured quantum
corrected string Bethe ansatz by demonstrating that the study of string sigma
model one-loop corrections 
for rational three-spin strings uniquely determines
the coefficients $\delta c_{r,s}$ to the same values as those found for the
$\mathfrak{sl}(2)$ sector in~\cite{Hernandez:2006tk}\footnote{While this manuscript
was beeing typed an interesting paper which gives another argument
in favour of the conjectured quantum string Bethe ansatz 
appeared~\cite{Arutyunov:2006iu}.}. 
Amazingly, despite the non-closure of sub-sectors of the quantum string
theory, a relatively simple dressing factor is capable of accounting
for at least the lowest order quantum effects. While the explicit form of
the one loop corrected dressing phase does not unambiguously explain the
three-loop discrepancy between gauge theory and strings it does leave
open the possibility that the discrepancy would be resolved by a full quantum
string computation. Earlier ideas to explain the discrepancy via 
so-called gauge theory wrapping interactions now seem to be ruled 
out~\cite{Ambjorn:2005wa,Rej:2005qt}. Obviously, we are in need for progress
on the quantization of the string theory. 
Recently, an alternative formulation of a quantum string Bethe ansatz 
for certain sub-sectors
has been proposed, based on an investigation of the $AdS_5 \times S^5$
superstring in light cone gauge~\cite{Frolov:2006cc}. Furthermore, 
interesting works 
addressing the string
quantization procedure via the study of various integrable sigma-models
in two dimensions have 
appeared~\cite{Mann:2005ab,Klose:2006dd,Gromov:2006dh}. With our universality
check we have provided evidence that any S-matrix resulting from such studies
should reproduce not only the classical string Bethe ansatz but also 
include the one-loop correction term.

\vspace*{1.0cm}
{\bf Acknowledgments}

\vspace*{0.3cm}
We thank N.\ Beisert for discussions.  
C.K.\ acknowledges the support of
ENRAGE (European Network on Random Geometry), a Marie Curie
Research Training Network supported by the European Community's
Sixth Framework Programme, network contract MRTN-CT-2004-005616.

\bibliographystyle{unsrt}
\bibliography{nonanalytic}
\end{document}